\documentclass[aps,twocolumn,showpacs,floatfix]{revtex4}
\usepackage{graphicx}
\newcommand{\be}{\begin{eqnarray}}
\newcommand{\ee}{\end{eqnarray}}
\def\beq{\begin{equation}}
\def\eeq{\end{equation}}

\begin{document}

\title{Anderson transition in a three dimensional kicked rotor}
\author{Jiao Wang}
\affiliation{Temasek Laboratories, National University of
Singapore, 117542 Singapore,} \affiliation{Beijing-Hong
Kong-Singapore Joint Center for Nonlinear and Complex Systems
(Singapore), National University of Singapore, 117542 Singapore}
\author{Antonio M. Garc\'{\i}a-Garc\'{\i}a}
\affiliation{Physics Department, Princeton University, Princeton,
New Jersey 08544, USA,} \affiliation{The Abdus Salam International
Centre for Theoretical Physics, P.O.B. 586, 34100 Trieste, Italy}

\begin{abstract}
We investigate Anderson localization in a three dimensional (3d)
kicked rotor. By a finite size scaling analysis we have identified a
mobility edge for a certain value of the kicking strength $k = k_c$.
For $k > k_c$ dynamical localization does not occur, all eigenstates
are delocalized and the spectral correlations are well described by
Wigner-Dyson statistics. This can be understood by mapping the
kicked rotor problem onto a 3d Anderson model (AM) where a band of
metallic states exists for sufficiently weak disorder. Around the
critical region $k \approx k_c$ we have carried out a detailed study
of the level statistics and quantum diffusion. In agreement with the
predictions of the one parameter scaling theory (OPT) and with
previous numerical simulations of a 3d AM at the transition, the
number variance is linear, level repulsion is still observed and
quantum diffusion is anomalous with $\langle p^2 \rangle \propto
t^{2/3}$. We note that in the 3d kicked rotor the dynamics is not
random but deterministic. In order to estimate the differences
between these two situations we have studied a 3d kicked rotor in
which the kinetic term of the associated evolution matrix is random.
A detailed numerical comparison shows that the differences between
the two cases are relatively small. However in the deterministic
case only a small set of irrational periods was used. A qualitative
analysis of a much larger set suggests that the deviations between
the random and the deterministic kicked rotor can be important for
certain choices of periods. Contrary to intuition correlations in
the deterministic case can either suppress or enhance Anderson
localization effects.

\end{abstract}

\maketitle

The quantum kicked rotor \cite{kick,reviz}, namely, a particle in
a circle periodically kicked (with period $T$) by a smooth
potential $V(q)=k\cos(q)$,
\begin{eqnarray}
{H}= \frac{p^2}{2}+V(q)\sum_{n}\delta(t-Tn),\label{kr}
\end{eqnarray}
has played a central role in the development of quantum chaos. This
is hardly surprising since, despite its simplicity, it has a very
rich and highly non-trivial dynamics. Moreover, it can be studied
 experimentally \cite{Raizen,otherexp} by using cold atoms
techniques.

The classical dynamics undergoes a gradual crossover from
quasi-integrable to fully chaotic as the kicking strength $k$ is
increased. Therefore it is possible to study in detail all types of
intermediate behavior between chaos and integrability. Moreover it
is specially suited for numerical simulations since the evolution
matrix in a basis of plane waves has a simple form. This was
important in the early days of quantum chaos when numerical
simulations in conservative 2d chaotic systems were quite expensive.

The quantum dynamics of the kicked rotor is also of interest. For
short time scales and $k$ sufficiently large quantum and classical
motion are diffusive. However for longer time scales numerical
simulations show \cite{kick} that quantum diffusion in momentum
space eventually stops, namely, $\lim_{t\to \infty} \langle p^2(t)
\rangle$ tends to a constant. By contrast classically $\langle
p^2(t) \rangle \propto t$ for all $t$. This rather counterintuitive
feature, usually referred to as dynamical localization \cite{reviz},
 was fully understood \cite{fishman} after mapping the kicked
rotor problem onto a 1d AM with a pseudo-random potential where
localization is well established. In our model the role of the mean
free path is played by the kicking strength $k$. The mapping onto a
1d AM is carried out in two steps: first one writes down the
evolution operator ${U}$ associated to Eq.(\ref{kr}), \be {U} =
e^{\frac{-i {\hat p}^2T}{2{\bar h}}} e^{-\frac{iV(\hat q)}{\bar h}}.
\ee In a basis of plane waves, $\hat{U}$ has matrix elements,
\begin{eqnarray}
U_{lm}= (-i)^{l-m}e^ {-i \frac{m^2 T \hbar}{2}} \
J_{l-m}\left(\frac{k}{\hbar}\right)\label{eq33}
\end{eqnarray}
where $J_n(k)$ is a Bessel function of first kind. Then a given
eigenstate of U with eigenvalue $\exp(-i\omega)$ can be transformed
into an effective Hamiltonian problem \cite{fishman},
\begin{eqnarray}
{\cal H} \psi_n= \epsilon_n \psi_n + \sum_m
F(m-n)\psi_m\label{ourmodel1}
\end{eqnarray}
where $\epsilon_n =\tan(\omega/2 -\tau n^2/4)$ and
\begin{eqnarray}
F(m-n) = - \frac{1}{2\pi}\int_{-\pi}^{-\pi}dq \tan(V(q)/2)e^{-iq(m-n)}\nonumber\\
\propto~\exp(-A|m-n|)~~~~~~~~~~~~~~~~
\end{eqnarray}
with $A$ a decreasing function of $k$ and $\tau = T/\hbar$. For
comparison we note that the standard 1d AM heavily used in
numerical and analytical studies of localization is given by, \be
\label{am} {\cal H} \psi_n= \epsilon_n \psi_n
+\psi_{n+1}+\psi_{n-1} \ee where $\epsilon_n$ are independent
random numbers extracted from a Gaussian or box distribution. The
quantum dynamics of the Hamiltonian Eq.(\ref{am}) is well
understood \cite{anderson,one,mac}: eigenstates are exponentially
localized and quantum diffusion eventually is arrested for any
amount of disorder.

We note that strictly speaking there are important differences
between Eq.(\ref{am}) and Eq.(\ref{ourmodel1}). For instance in
Eq.(\ref{ourmodel1}) the diagonal disorder $\epsilon_n$ is
deterministic and the hopping is short range but it is not
restricted to nearest neighbors as in Eq.(\ref{am}). The latter is
not really relevant as it is known \cite{mirfyo} that exponential
localization for all energies and disorder persists provided that
the hopping term decays faster than $~1/|n-m|$. Potentials in
Eq.(\ref{kr}) leading to a decay of $F(m-n)\sim 1/|n-m|$ in
Eq.(\ref{ourmodel1}) induce a metal-insulator transition even in
1d \cite{usprl,mirfyo}.

The non-randomness of the diagonal disorder in
Eq.(\ref{ourmodel1}) is, at least  superficially, a more serious
deviations from Eq.(\ref{am}). However, in Ref. \cite{fishcorr} it
was showed that provided $\tau \gg 1$ is irrational with a
sufficiently poor approximation by a rational number, the
potential was ergodic and to a great extent not very different
from a true random potential. This seems to be related to the fact
that in the $n \to \infty$ limit all derivatives of $\epsilon_n$
are not bounded. For the sake of completeness we mention that a)
it was not until recently \cite{bour1} that a rigorous proof of
localization for the original model Eq.(\ref{kr}), not involving a
mapping onto an AM, is available; and b) localization in more
general time dependent potentials has been recently studied in the
mathematical literature. The main conclusion \cite{bour2} is that
if the potential is periodic and smooth in time, dynamical
localization will generally occur.

A natural question to ask is to what extent these results hold in
higher dimensions. One of the more striking predictions of
localization theory \cite{one,mac}, corroborated by numerical
simulations, is that eigenstates of the 2d version of Eq.(\ref{am})
are still exponential localized for any disorder and energy. As in
the 1d case, quantum diffusion eventually stops as well. However in
more than 2d a metal insulator transition occurs for a critical
value of the disorder. The critical disorder increases with the
spatial dimensionality of the system \cite{one,anderson,mac}. It is
thus highly desirable to determine a) whether it is possible to
extend the above mapping to higher dimensions, and b) whether the
resulting extension of Eq.(\ref{ourmodel1}) still follows the
predictions of localization theory despite the potential being
non-random.

We note that if the two models are still similar higher dimensional
kicked rotors could be used to study experimentally the Anderson
transition by using cold atoms techniques. This is specially
relevant as experimental detection of localization has proved to be
quite elusive. Moreover it is also important to understand in detail
how deviations from an ideal random distribution affects the quantum
dynamics of the system. After all any realistic random potential
will always present some sort of deviation from a given ideal random
distribution.

In fact there are already in the literature different propositions
to map a generalization of Eq.(\ref{kr}) onto a higher dimensional
AM. In Ref. \cite{mulfreq} a multifrequency, time dependent
potential \be V(q)= k\cos( q +\omega_1 t + \omega_2 t) \ee was
mapped onto a 3d version of Eq.(\ref{ourmodel1}); namely, an AM with
short range hopping and diagonal disorder \be
\label{del}\epsilon_{n,n_1,n_2} \sim \tan(\tau n^2 + \omega_1n_1
+\omega_2n_2). \ee with $n,n_1,n_2$ integers numbers. A numerical
analysis of the quantum diffusion \cite{mulfreq} showed that, as in
a 3d AM with a truly random potential, the system avoids dynamical
localization for $k > k_c$. Recently it has been possible to model
this generalized kicked rotor by using cold atoms in an optical
lattice \cite{delande}. In agreement with previous results no
dynamical localization was observed for $k
>k_c$.

However we note that in the $n_1$ and $n_2$ direction the resulting
diagonal disorder is quasi-periodic rather than quasi-random. It is
thus unclear whether this type of kicked rotor can be used as an
effective model to study a metal-insulator transition induced by
disorder. Indeed the value of the critical exponent \cite{borgo}
controlling the divergence of the localization length in
generalizations of Eq.(\ref{del}) with $d \geq  3$ frequencies
$\omega_1 \ldots \omega_d$ sharply disagrees with numerical results
\cite{numand,mac} of
the d-dimensional equivalent of Eq.(\ref{am}).\\

Another generalized kicked rotor with potential, \be V(q_1,q_2) = k
\cos(q_1)\cos(q_2) \ee was studied in Ref. \cite{doron}. In this
case the mapping is onto a 2d AM with short range hopping and
diagonal disorder, \be \epsilon_{n_1,n_2}\sim \tan(\tau_1 n_1^2+
\tau_2 n_2^2). \ee with $n_1,n_2$ integers and $\tau_1,\tau_2$
irrationals. This diagonal pseudo-disorder is a natural extension of
the one dimensional case Eq.(\ref{ourmodel1}) and consequently
shares many of the properties of a true random potential. For
instance, the dynamical localization occurs for any value of the
kicking strength, and the localization length increases
exponentially with $k$.

In this paper we study a 3d generalization of the above model,
namely, a 3d kicked rotor with \be V(q_1,q_2,q_3) = k
\cos(q_1)\cos(q_2)\cos(q_3). \ee It will be shown that this model
maps onto a 3d version of Eq.(\ref{ourmodel1}) with short range
hoping and an on-site potential that shares many quasi-random
features of the 1d and 2d cases.

By a careful finite size scaling analysis it will shown that this
model undergoes a metal insulator transition at a certain kicking
strength $k = k_c$. For technical reasons we have investigated in
detail only a small sets of irrational periods $\{\tau_i \}$ with
$i = 1, 2, 3$. Within this set the numerical value of $k_c$
depends weakly on the $\{\tau_i \}$.  Dynamical localization
consequently only occurs for $k < k_c$. We then study in detail
level statistics and quantum diffusion in the critical region
$k\approx k_c$. In agreement with previous numerical results of
the level statistics of the 3d AM at the transition, we have found
that a) the spectral rigidity is linear; b) as in a metal, level
repulsion is still observed and, c) as in an insulator, the
asymptotic decay of the level spacing distribution is exponential.
Moreover, in agreement with the the one parameter scaling theory
(OPT), the diffusion is anomalous at the transition with $\langle
p^2 \rangle \propto t^{2/3}$.

Finally we investigate to what extent Anderson localization is
affected by the pseudo random nature of the potential. In order to
proceed we compare the results of previous sections with those
coming from a 3d kicked rotor in which the phase of the kinetic term
of the evolution matrix is random. The main differences between the
random and pseudo random cases are as follow: a) the critical
kicking strength $k_c$ is slightly larger $(k_c \sim 2.4$ versus
$k_c \sim 2.3)$ in the random case; b) the slope of the spectral
rigidity is smaller in the random case and c) statistical
fluctuations are much stronger in the pseudo random case.  These
results are consistent with the heuristic picture that localization
will be stronger the more random/uncorrelated the potential is.
However a qualitative analysis of a much larger set of $\{\tau_i \}$
suggests that the deviations between the random and the
deterministic kicked rotor can be important for certain choices of
periods. Contrary to intuition certain choices of $\{\tau_i \}$ can
enhance Anderson localization effects and lead to a larger $k_c$
than in the random case.

We start with a brief review of the features expected at the
metal-insulator transition according to numerical simulations
\cite{numand} and OPT \cite{one}.

%%%%%%%%%%%%%%%%%%%%%%%%%%%%%%%%%%%%%%%%%%%%%%%%%%%%%%%%%%%%%%%%%%%%%%%%%%%%%%%%%%%%%%%%%%%%%%%%%%%fig1
\begin{figure*}
\includegraphics[width=.95\columnwidth,clip]{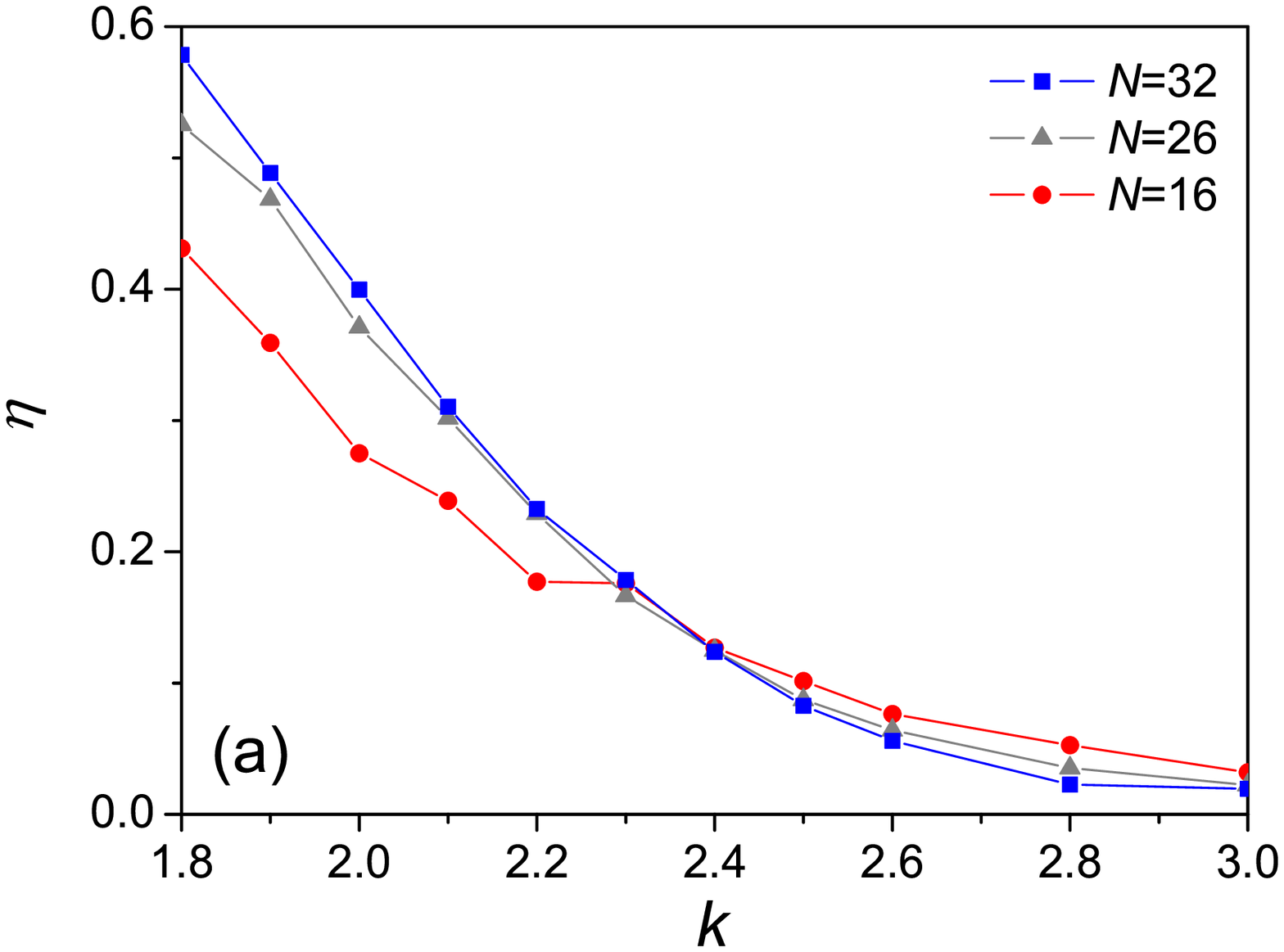}
\includegraphics[width=.95\columnwidth,clip]{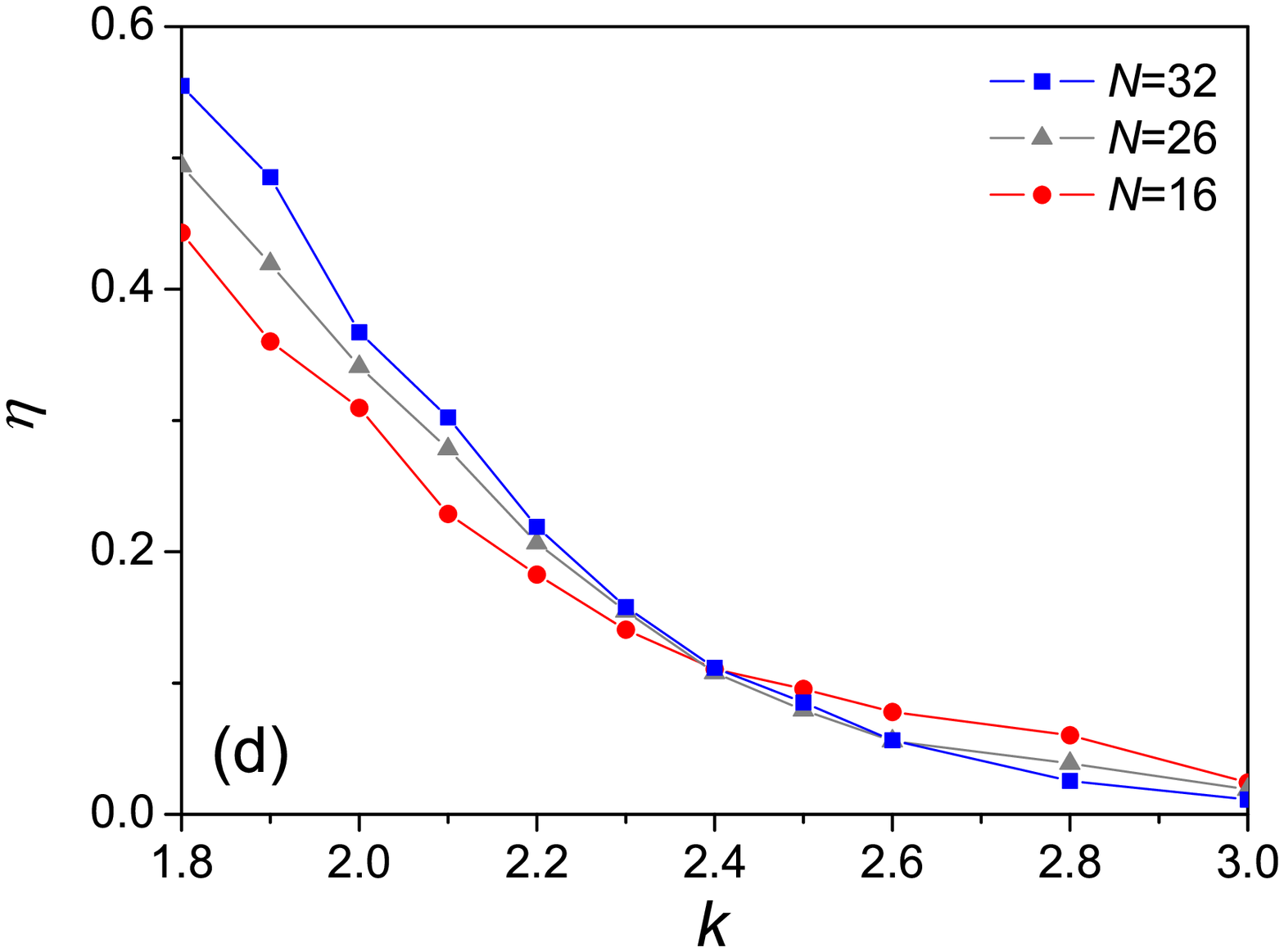}
\includegraphics[width=.95\columnwidth,clip]{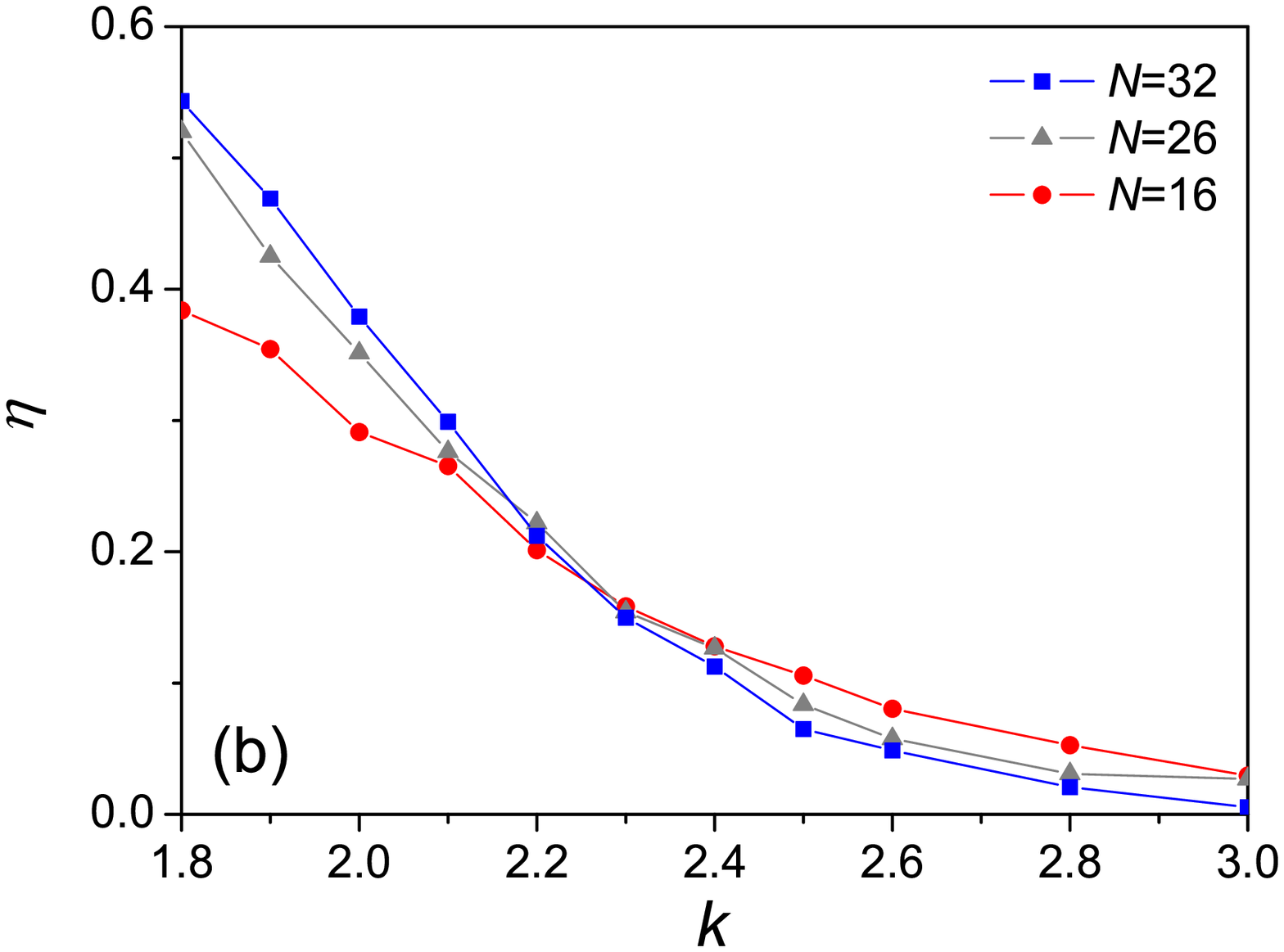}
\includegraphics[width=.95\columnwidth,clip]{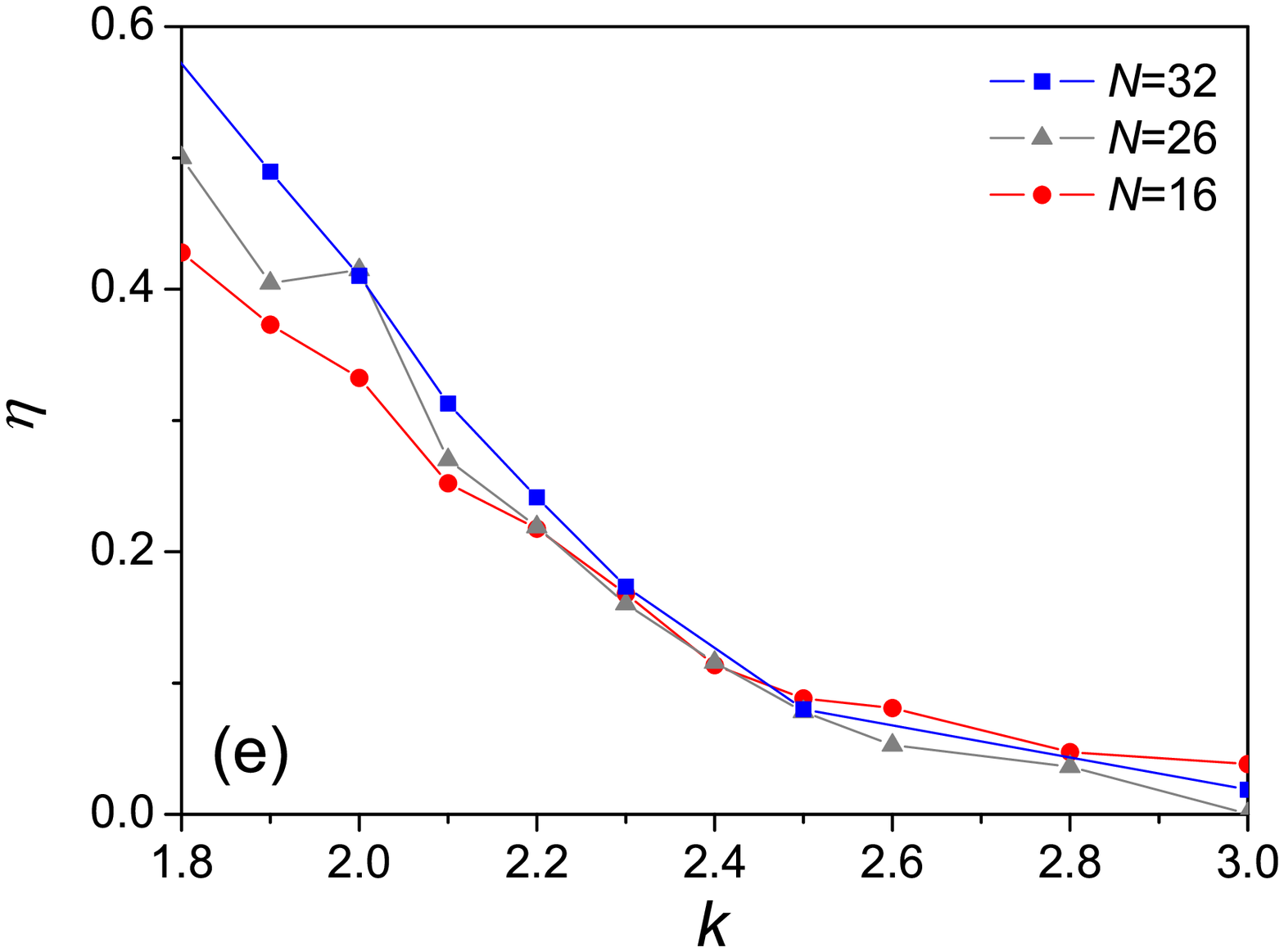}
\includegraphics[width=.95\columnwidth,clip]{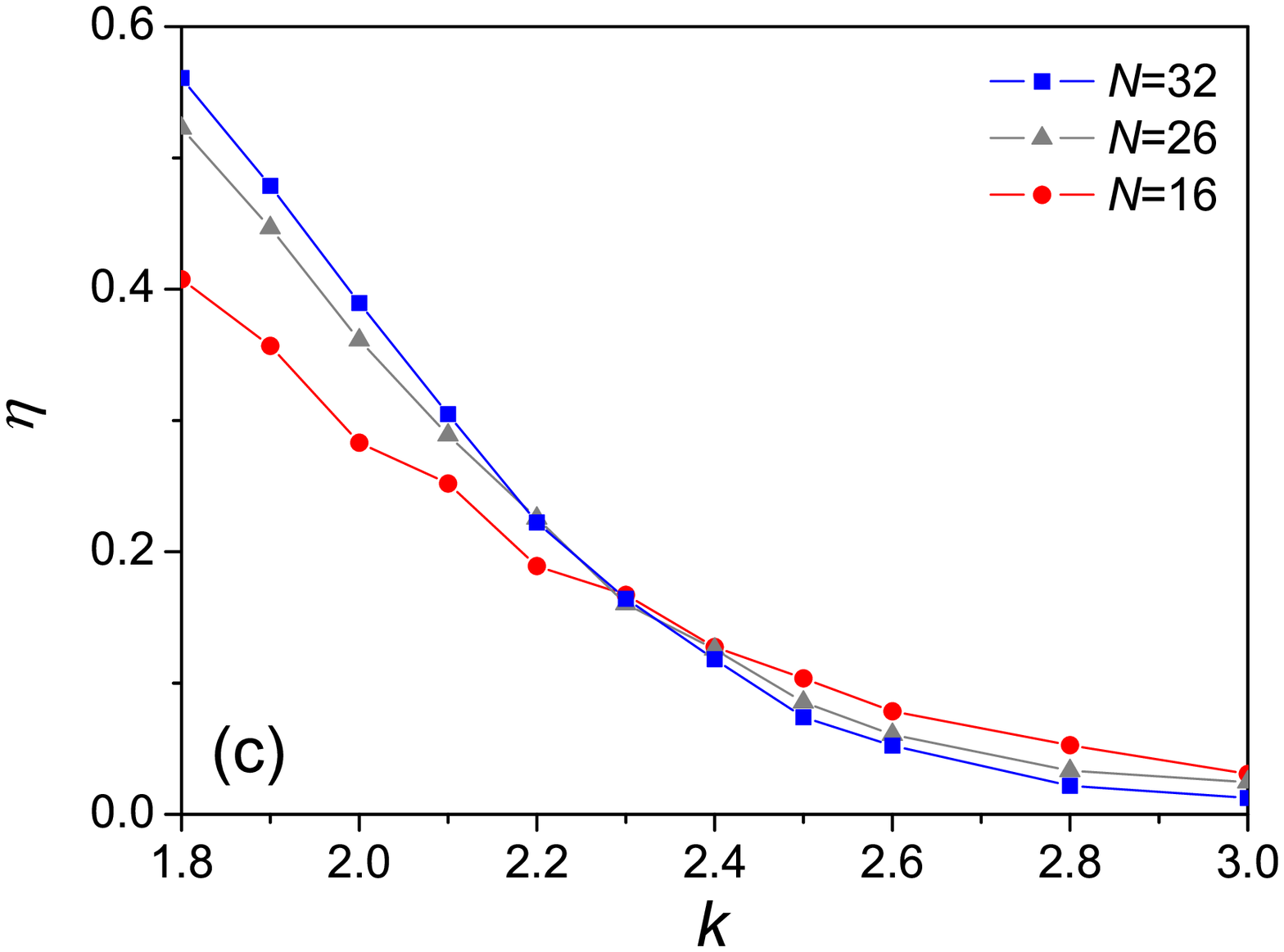}
\includegraphics[width=.95\columnwidth,clip]{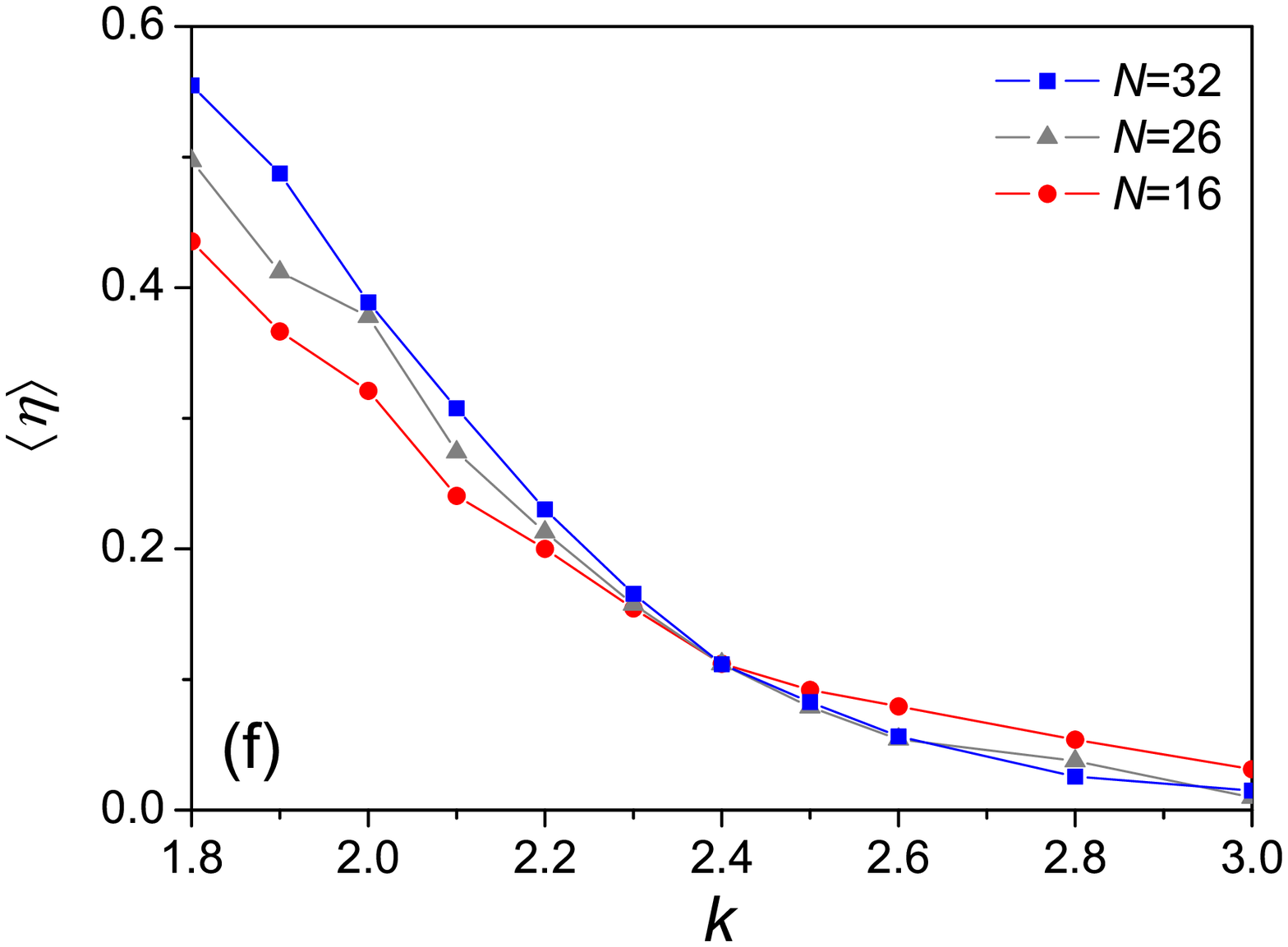}
\caption{Scaling variable $\eta$ given by Eq. (\ref{scaling}) as a
function of $k$ for different system sizes $N$. (a) and (b) are
the results for two different sets of $\{\tau_i\}$ generated with
Method 1 (see text), (c) are the averaged results of (a) and (b).
The finite size scaling analysis suggests that a metal-insulator
transition occurs at $k_c=2.27\pm0.07$ with a critical exponent
$\nu=1.61\pm0.25$. Plots (d), (e) and (f) are the equivalent of
plots (a), (b) and (c) but with $\{\tau_i \}$ obtained by Method 2
instead. In this case we obtain $k_c=2.35\pm0.1$ with
$\nu=1.67\pm0.27$.} \label{fig1}
\end{figure*}
%%%%%%%%%%%%%%%%%%%%%%%%%%%%%%%%%%%%%%%%%%%%%%%%%%%%%%%%%%%%%%%%%%%%%%%%%%%%%%%%%%%%%%%%%%%%%%%%%%%fig1

\subsection{The one parameter scaling theory and properties of the 3d metal-insulator transition}

In the context of disordered systems the one parameter scaling
theory (OPT) \cite{one} provides a valuable framework to
understand localization effects. In two and lower dimensions it
predicts the exponential localization of the eigenstates and the
arrest of diffusion for any amount of disorder \cite{one}. In
three and higher dimension a metal insulator transition is
expected for a critical amount of disorder. Thus for disorder
below the critical one the wavefunctions are extended through the
sample. In the opposite limit, wave functions are exponentially
localized as in $d \leq 2$.

A key concept in this theory is the dimensionless conductance $g$
introduced  by Thouless \cite{thouless2}. It is defined as $g =
E_T/\Delta$ where, $E_T$, the Thouless energy, is an energy scale
related to the diffusion time to cross the sample and $\Delta
\propto 1/L^d$ is the mean level spacing. In the diffusive limit
$E_T \propto 1/L^2$ and therefore $g \propto L^{d-2}$. On the other
hand if the the particle is exponentially localized, $g \propto
e^{-L/\xi}$ where $\xi$ is the localization length and $L$ is the
system linear size.

%The change of the dimensionless conductance with the system size
%is an indicator of localization. Qualitatively, in the $L \to
%\infty$ limit,  $g \to \infty$ in a metal and  $g \to 0$ in an
%insulator. In order to proceed it is useful to define \be \beta(g)
%= \frac{\partial \log g(L)}{\partial \log L} \ee which describes
%the running of $g$ with the system size.

%From the above definitions, $\beta(g) = d-2 > 0$ in a metal
%(without quantum corrections) and $\beta(g)= \log(g) <0$ in an
%insulator.

With this simple input it is possible to analyze localization
effects in 3d: For a metal $\lim_{L \to \infty} g(L) \to \infty$.
For an insulator $\lim_{L \to \infty} g(L) \to 0 $. Since $g(L)$ is
continuous and monotonous, the metal insulator transition is
characterized by a size independent conductance $g = g_c$. Since
$g_c$ does not depend on the system size and $\Delta \propto 1/L^3$,
the Thouless energy must scale as $E_T \propto 1/L^3$. This only can
happen if the diffusion is anomalous at the transition with \be
\langle r^2\rangle \sim t^{2/3}. \ee Anomalous diffusion is one of
the main predictions of the OPT about the metal insulator
transition.

Additional information about the transition comes mainly from
numerical calculations \cite{numand}. Level statistics
\cite{kravtsov,sko} are intermediate between those of a metal and
those of an insulator. For instance level repulsion is observed as
in a disordered metal. By contrast long range spectral
correlations such as the spectral rigidity $\Delta_3(\Lambda)$ are
closer to that of an insulator \cite{chi,antprl}. For $\Lambda \gg
1$, \be \Delta_{3}(\Lambda)=\frac{2}{\Lambda^4}\int_{0}^{\Lambda}
(\Lambda^3-2\Lambda^2x+x^3)\Sigma^{2}(x)dx \sim \chi \Lambda/15
\nonumber \ee where $\chi  < 1$ at the metal insulator transition
and $\chi = 1$ for an insulator, $\Lambda$ is a spectral window
containing $\Lambda$ eigenvalues in units of the mean level
spacing $\Delta$, $\Sigma^{2}(\Lambda)=\langle N_\Lambda^2 \rangle
- \langle N_\Lambda \rangle^2$ is the number variance ($N_\Lambda$
is the number of eigenvalues in an interval of length $\Lambda$).

The critical exponent controlling the divergence of the localization
length at the transition is supposed to be a decreasing function of
the space dimensionality. In 3d it is approximately given by $\nu
\approx 3/2$\cite{numand}. Recently one of us showed \cite{antprl}
that some of these features can be obtained analytically by
combining the OPT with the self-consistent theory of localization
\cite{self}.

%In this paper we explore to what extent these results also hold in
%our 3d kicked rotor Eq.(\ref{3dkr}). In the next section we carry
%out a finite size scaling analysis which clearly shows that a metal
%insulator transition occurs in our model for $k=k_c \approx 2.4$.
%Then in the following sections we study level statistics, quantum
%diffusion and critical exponents. Finally we compare our previous
%results with a generalized 3d kicked rotor in which the kinetic term
%of the associated evolution matrix is made random by hand. Our
%motivation is to examine quantitatively the effect of
%pseudo-randomness on Anderson localization effects.

\section{Calculation of the critical strength $k_c$ and critical exponent $\nu$}

We study the quantum dynamics of a 3d kicked rotor with a smooth
potential: \be \label{3dkr} {\cal H}= \frac{1} 2 (\tau_1 p_1^2
+\tau_2 p_2^2+\tau_3 p_3^2)~~~~~~~\nonumber\\
+V(q_1,q_2,q_3)\sum_n\delta(t -n) \ee with \be V(q_1,q_2,
q_3)=k\cos(q_1)\cos(q_2)\cos(q_3), \ee  $T =\hbar \equiv 1$,
$\tau_1,\tau_2,\tau_3$ are irrational numbers not close to any
rational multiple of $\pi$. It is important to choose $\{\tau_i\}$
in such a way that correlations caused by the proximity to a
rational number are suppressed. However it is hard to
quantitatively assess how the specific choice of $\{\tau_i\}$ will
influence Anderson localization effects. In 3d this is rather a
number theory problem which is beyond the scope of this paper.
Here in order to qualitatively address this issue we have employed
two different methods to generate a given set of $\{\tau_i \}$:

{\it Method 1:} We set $\tau_1 = \alpha/\lambda^2$, $\tau_2 =
\alpha/\lambda$ and $\tau_3 = \alpha$ where $\lambda = 1.3247....$
is the real root of the equation $\lambda^3 -\lambda -1 = 0$, and
$\alpha$ is a random number extracted from a box distribution with
support $(10,20)$. In this way $\tau_i \gg 1$. This is necessary
condition in order to have pseudo-random features. A given
$\alpha$ is only accepted if the resulting continuous fraction
form of $\tau_1/\pi, \tau_2/\pi$ and $\tau_3/\pi$ specified by a
sequence of integers, say, $[a_1, a_2, ...]$ is such that the
maximum value of $|a_i|$ is smaller than 12. This is enough to
prevent the system to be close to a quantum resonance. We note
that the smaller these integers $a_i$ are the more irrational are
$\tau_i/\pi$.

{\it Method 2:} We set $\tau_1 = \alpha/\sigma$, $\tau_2 = \alpha$
and $\tau_3 = \sigma\alpha$ where $\sigma \approx 1.618$ is the
golden mean. Then we follow the Method $1$.

%{\bf do you know why these two methods are different?, for
%instance, why we did not just set $\tau_i=\alpha$ and then pick up
%those $\alpha's$ which are good irrational}.

%In fig \ref{fig6} we plot the correlations of the kinetic term of
%the evolution operator $C(i_1,i_2,i_3 =0)$ {\bf add definition}
%(see below) by using these two methods. For a white noise we would
%see a peak for $i_1=i_2=0$ only. However in our case we see for
%both methods  a much more intricate pattern.

In this section we carry out a finite size scaling analysis in order
to determine the impact of this type of pseudo randomness on
Anderson localization effects. We will show that the system
undergoes a metal insulator transition at $k = k_c = 2.27 \pm 0.07$
if Method $1$ is used and $k = k_c = 2.35 \pm 0.1$ if Method $2$ is
used. By contrast if the phase of the kinetic term of the evolution
matrix is completely random (see last section) $k_c = 2.40 \pm
0.05$. It thus clear that the pseudo random nature of the $\{\tau_i
\}$ have a sizable but small effect on the localization transition.
%Although it cannot avoid the eventual transition to an insulator it
%decreases the critical kicking strength.
% Qualitatively it is expected that in
%pseudo random systems localization effects are weaker and
%consequently the value of $k_c$, which is inversely proportional to
%disorder, is smaller.
We shall see in the last section that for certain choices  of
$\{\tau_i \}$ these differences can be much larger and likely lead
to $k_c$ quite different from the one corresponding to the random
case.

 In order to proceed we evaluate the quantum evolution
operator $U$. After a period $T \equiv 1$ (we remember that $\hbar
\equiv 1$), an initial state $\psi_0$ evolves to \be \psi(1) =
{U}\psi_0 = e^{\frac{-i {\hat p}^2}{2}} e^{-{iV(\hat q)}}\psi_0, \ee
where $\hat p$ and $\hat q$ stand for the usual momentum and
position operator. We then express the evolution operator $U$ in a
basis of the momentum eigenstates,
%by using a basis of plane
%waves $| n \rangle = \frac{e^{in \theta}}{\sqrt{2\pi}}$
%(where $n,m=1,\ldots N$)
%which are the eigenvalues of the momentum operator.
%The resulting matrix  is Unitary
%exclusively in the thermodynamic limit.
\be \label{uni} \langle {\bf m}| { U}| {\bf n} \rangle =
\frac{1}{N}e^{-i(\tau_1 n_1^2+\tau_2 n_2^2+\tau_3 n_3^2)/2}
\sum_{\bf l}e^{i\phi({\bf l},{\bf m},{\bf n})} \ee where $\langle
{\bf q}| {\bf n} \rangle = \frac {e^{i {\bf n}{\bf
q}}}{(2\pi)^{3/2}}$ with ${\bf q}=(q_1,q_2,q_3)$, ${\bf
n}=(n_1,n_2,n_3)$, $\phi({\bf l},{\bf m}, {\bf n})= 2\pi ({\bf
l}+{\bf \theta})({\bf m}- {\bf n})/N-V(2\pi ({\bf
l}+{\bf\theta})/N)$, ${\bf l} = (l_1,l_2,l_3)$, $l_i = -N/2,\ldots ,
N/2-1$. Here $N^3$ specifies the dimension of the Hilbert space. We
restrict ourselves to even $N$ throughout this work. The parameter
${\bf \theta}$ depends on the boundary conditions $0 \le \theta_i\le
1$. We set $\bf{\theta}=\bf{0}$ that corresponds to the periodic
boundary conditions.

%Besides, $\hbar=1$ and $T=1$ throughout this work.
%{\bf I remember you computed explicitly the evolution matrix in
%terms of integrals of Bessel functions. Could you maybe include it
%or simply replace Eq.(\ref{uni})?}

The eigenvalues and eigenvectors of $U$ can now be computed by
using standard diagonalization techniques.
%However we found more efficient to compute them by evolving a quantum state
%$|\psi(0)\rangle$ and performing the Fourier transform of
%$\langle\psi(t)|\psi(0)\rangle$. {\bf do you want to add something
%about the numerical calculation, how many set of $\tau$'s did you
%use...?, why we discard small volumes...}
The eigenvalues of the evolution matrix are the starting point to
carry out the finite size scaling analysis.

Finally we note that following the procedure of Ref. \cite{fishman}
it is possible to map a given eigenstate of U with eigenphase
$e^{i\omega}$ onto a 3d AM, \be \label{ourmodel2} {\cal H} \psi_{\bf
n}= \epsilon_{\bf n} \psi_{\bf n} + \sum_{\bf m} F({\bf m}-{\bf
n})\psi_{\bf m} \ee where
%${\vec n} = (n_1,n_2,n_2)$,
\be \label{enoran}\epsilon_{\bf n} = \tan(\omega/2-\tau_1 n_1^2/4 -
\tau_2 n_2^2/4 -\tau_3 n_2^2/4) \ee and \be F({\bf m}- {\bf n}) =-
\frac{1}{(2\pi)^3}\int_{-\pi}^{-\pi}d^3q \tan(V({\bf q})/2)e^{-i{\bf
q}({\bf m}-{\bf n})}\nonumber\\ \propto \exp(-A|{\bf m}-{\bf
n}|)~~~~~~~~~~~~~~~~~~~ \nonumber \ee with $A$ a decreasing function
of the kicking strength.
%{\bf could you please check that and modify them if it is
%not right}.
The form of the diagonal disorder and hopping is similar to the 1d
case. We thus expect that the quantum dynamics of
Eq.(\ref{ourmodel2}) is qualitatively similar to that of a 3d AM
with uncorrelated disorder and hopping restricted to nearest
neighbors. However deviations are expected since the potential
Eq.(\ref{enoran}) have strong correlations (see Fig. \ref{fig5}).

\subsection{Finite size scaling method}
Our first task is to find out the critical $k_c$ and the critical
exponent $\nu$ associated with the divergence of the localization
length at the transition. In order to proceed we determine $k_c$
by using the finite size scaling method \cite{sko}. First we
evaluate a certain spectral correlator for different sizes $N$ and
disorder strengths $k$. Then we locate the transition by finding
the kicking strength  $k_c$ such that the chosen scaling variable
becomes size-independent. In our case we investigate the level
spacing distribution $P(s)$ \cite{mehta} (probability of finding
two neighboring eigenvalues at a distance $s = (\epsilon_{i+1} -
\epsilon_{i})/\Delta $, with $\Delta$ being the local mean level
spacing). In order to avoid any dependence on bin size, the
scaling behavior of $P(s)$ is examined through the following
function of its variance\cite{C99}
\begin{equation}
\label{scaling} \eta (N,W) = [{\rm var} (s)- {\rm var_{WD}}\,] /
[\,{\rm var_{P}}-{\rm var_{WD}}].\label{eta}
\end{equation}
In Eq. (\ref{eta}) ${\rm var} (s)=\langle s^2\rangle-\langle s
\rangle^2$, where $\langle \dots \rangle$ denotes spectral
averaging over a single set of $\{\tau_i\}$. ${\rm var_{WD}}
\approx 0.286$ and ${\rm var_P}=1$ are the variances of
Wigner-Dyson (metal) and Poisson (insulator) statistics
\cite{mehta}, respectively. Hence $\eta = 1 (0)$ for an insulator
(metal). Any other intermediate value of $\eta$ in the $N \to
\infty$ limit is an indication of a metal insulator transition.

%%%%%%%%%%%%%%%%%%%%%%%%%%%%%%%%%%%%%%%%%%%%%%%%%%%%%%%%%%%%%%%%%%%%%%%%%%%%%%%%%%%%%%%%%%%%%%%%%%%fig 2
\begin{figure}
\includegraphics[width=0.95\columnwidth,clip]{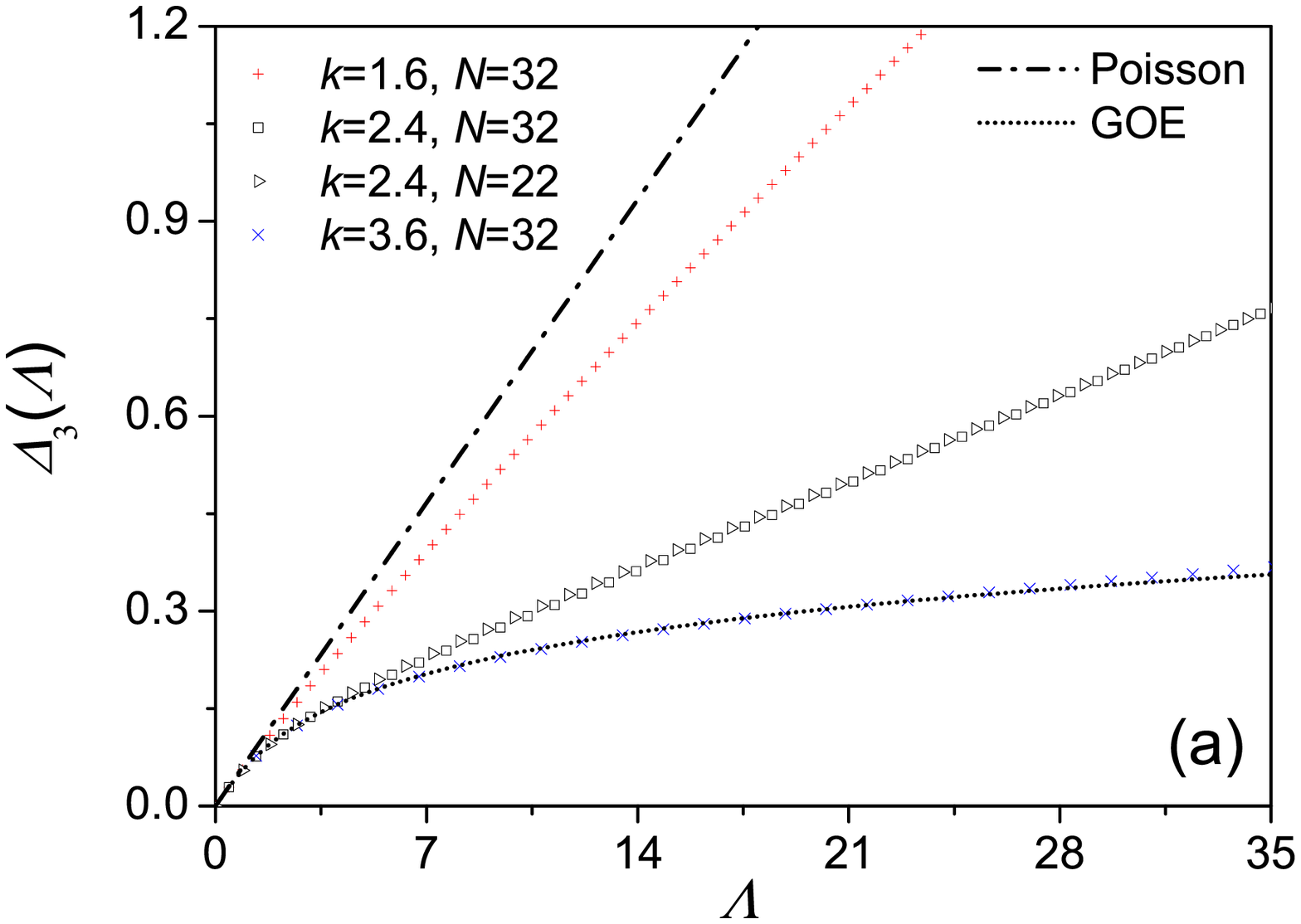}
\includegraphics[width=0.95\columnwidth,clip]{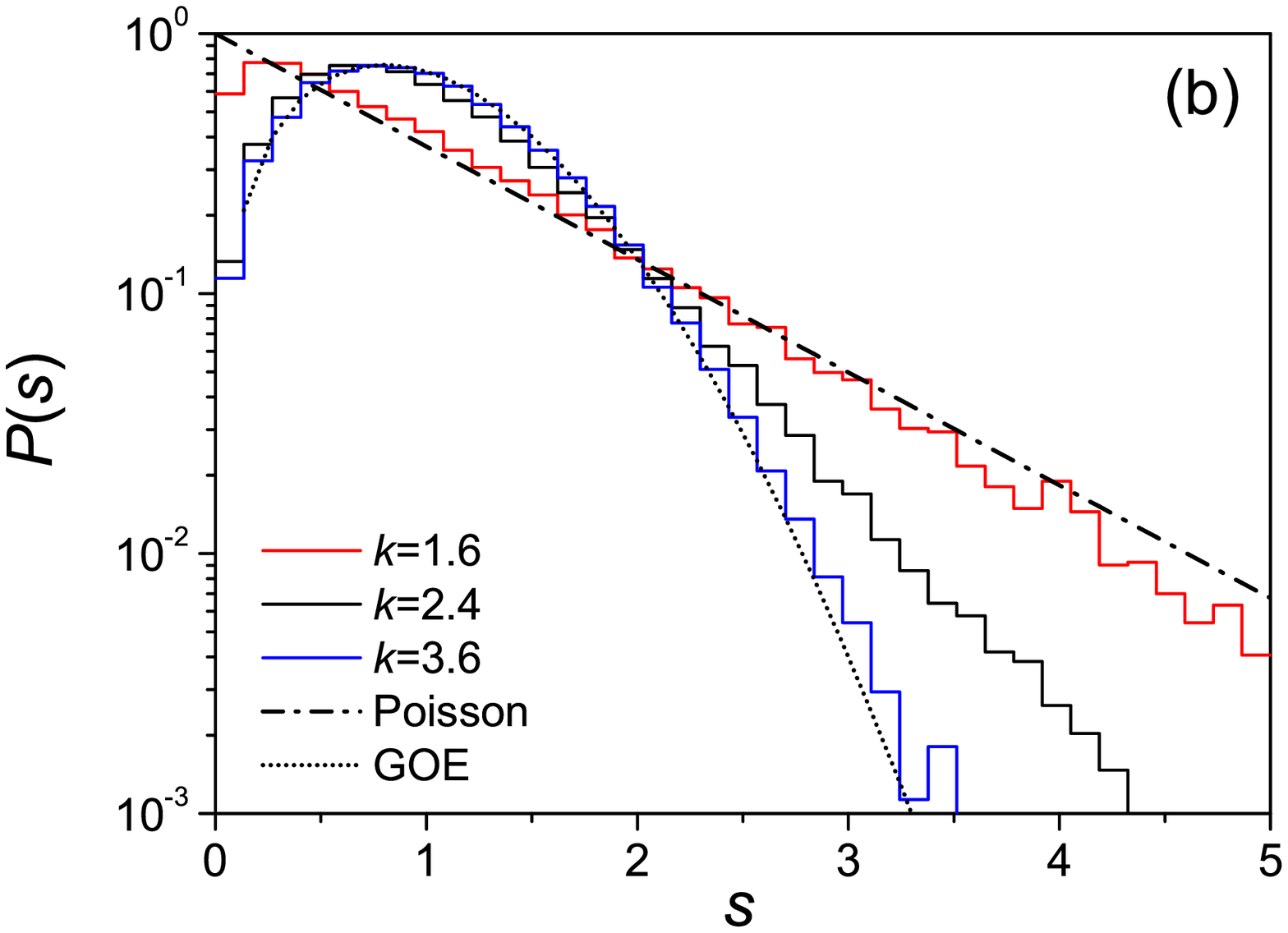}
\caption{The spectral rigidity $\Delta_3(\Lambda)$ and level spacing
distribution $P(s)$ for three different kicking values, above, below
and close to the transition $k_c \approx 2.4$ in the 3d kicked rotor
Eq. (\ref{3dkr}) with $\{\tau_i\}$ the same as in Fig.1(a). A
transition from Poisson to Wigner-Dyson statistics for a system with
time reversal invariance (GOE) is clearly observed as the kicking
strength crosses the critical value $k \approx k_c$ from below. For
$k \approx k_c$ the level statistics have all the signatures of a
metal-insulator transition such as a) an asymptotically linear
$\Delta_3(\Lambda)$, b) level repulsion for $s \to 0$ as in a metal
($\lim_{s \to 0}P(s) = 0$) and c) exponential decay of $P(s)$ for $s
\gg 1 $ as in an insulator. In agreement with the OPT, level
statistics are scale invariant for $k \approx k_c$.} \label{fig2}
\end{figure}
%%%%%%%%%%%%%%%%%%%%%%%%%%%%%%%%%%%%%%%%%%%%%%%%%%%%%%%%%%%%%%%%%%%%%%%%%%%%%%%%%%%%%%%%%%%%%%%%%%%fig 2

In Fig. 1 we plot the $k$ dependence of $\eta$ for different
system sizes. The critical disorder $k = k_c$ signaling the
Anderson transition corresponds to the point for which $\eta$ is
independent of $N$. For a smaller (larger) $k$, $\eta$ tends to
the insulator (metal) prediction. For a precise determination of
the critical $k_c$ and the critical exponent $\nu$ we look at the
correlation length near $k_c$,
\begin{equation}
\xi(k) = \xi_0 |k-k_c|^{-\nu}\;, \label{xiw}
\end{equation}
where $\xi_0$ is a constant. The numerical values of $k_c$ and
$\nu$ are obtained by expressing $\eta(N,k)=f[N/\xi(k)]$ and then
performing an expansion around the critical point
\begin{equation}
\eta(N,k) = \eta_c+ \sum_{n}C_n(k-k_c)^nN^{n/\nu}\;. \label{fit}
\label{alfa1}
\end{equation}
In practice, we have truncated the series at $n=2$ and then we have
performed a statistical analysis of the data with the
Levenberg-Marquardt method for nonlinear least-squares models. The
most likely fit is determined by minimizing the $\chi^2$ statistics
of the fitting function (\ref{fit}). We found  $k_c =2.27 \pm 0.07$
with $\nu=1.61\pm 0.25$ for Method $1$ and $k_c =2.35 \pm 0.1$ with
$\nu=1.67\pm 0.27$ for Method $2$. The numerical value of $\nu$ is
higher but quite close to the one found in the 3d AM at the
transition $\nu \approx 1.5$. We stress that these results were
obtained with only the data of Fig. 1, namely, an average over only
two sets of $\{\tau_i \}$ for each method.

 These findings are consistent with
the heuristic picture that correlations make the potential less
random and consequently weaken localization effects. We note that
this conclusion (see next section) might be misleading due to the
small number of $\{\tau_i \}$ utilized. Indeed in the final section
we shall see that our results are typical, namely, for the majority
of $\{\tau_i \}$ we shall get similar results. However there are
some choices of irrational $\{\tau_i \}$ that can lead to larger
deviations from the results presented in this section.

\section{Level statistics}
As was mentioned previously the analysis of level statistics is a
powerful tool to investigate the metal insulator transition. At the
transition level statistics are intermediate between that of a metal
and that of a insulator. For instance, a) $P(s) \sim s$ for $s \ll
1$, a typical feature of a disordered metal, is still observed at
the transition; b) $\Delta_3(\Lambda) = \chi \Lambda/15$ for
$\Lambda \gg 1$ with $\chi < 1$
\cite{chi,kravtsov,antprl,numand,moun}. This is similar to the
result for an insulator $\Delta_3(\Lambda) = \Lambda/15$; c)
likewise $P(s) \propto e^{-As}$ with $s \gg 1$ and $A > 1$. For an
insulator $P(s) = e^{-s}$.

We explore to what extent these features are also observed in our 3d
kicked rotor. As is shown in Fig. \ref{fig2}, qualitatively the
level statistics of the eigenphases of the evolution matrix at  $k
\approx k_c$ are strikingly similar to those of a 3d AM at the
transition. The spectral rigidity $\Delta_3(\Lambda)$ is linear,
there is level repulsion for $s \ll 1$, the spectrum is scale
invariant and $P(s)$ decays exponentially for $s  \gg 1$.  The slope
of $\Delta_3(\Lambda)$, $\chi \approx 0.29$ in our case, is larger
than the one corresponding to a random kinetic term (see next
section). This suggests that the quasi-random nature of our model
only modifies quantitatively, non-qualitatively, the transition. As
was expected, level statistics in the limits $k \gg (\ll) k_c$ are
described by Wigner-Dyson (Poisson) statistics typical of a metal
(insulator). We note that for systems with time reversal invariance
the Wigner-Dyson statistics is usually defined by the spectral
properties  of the Gaussian Orthogonal Ensemble of random matrices
(GOE).

%%%%%%%%%%%%%%%%%%%%%%%%%%%%%%%%%%%%%%%%%%%%%%%%%%%%%%%%%%%%%%%%%%%%%%%%%%%%%%%%%%%%%%%%%%%%%%%%%%%fig 3
\begin{figure}
\includegraphics[width=0.95\columnwidth,clip]{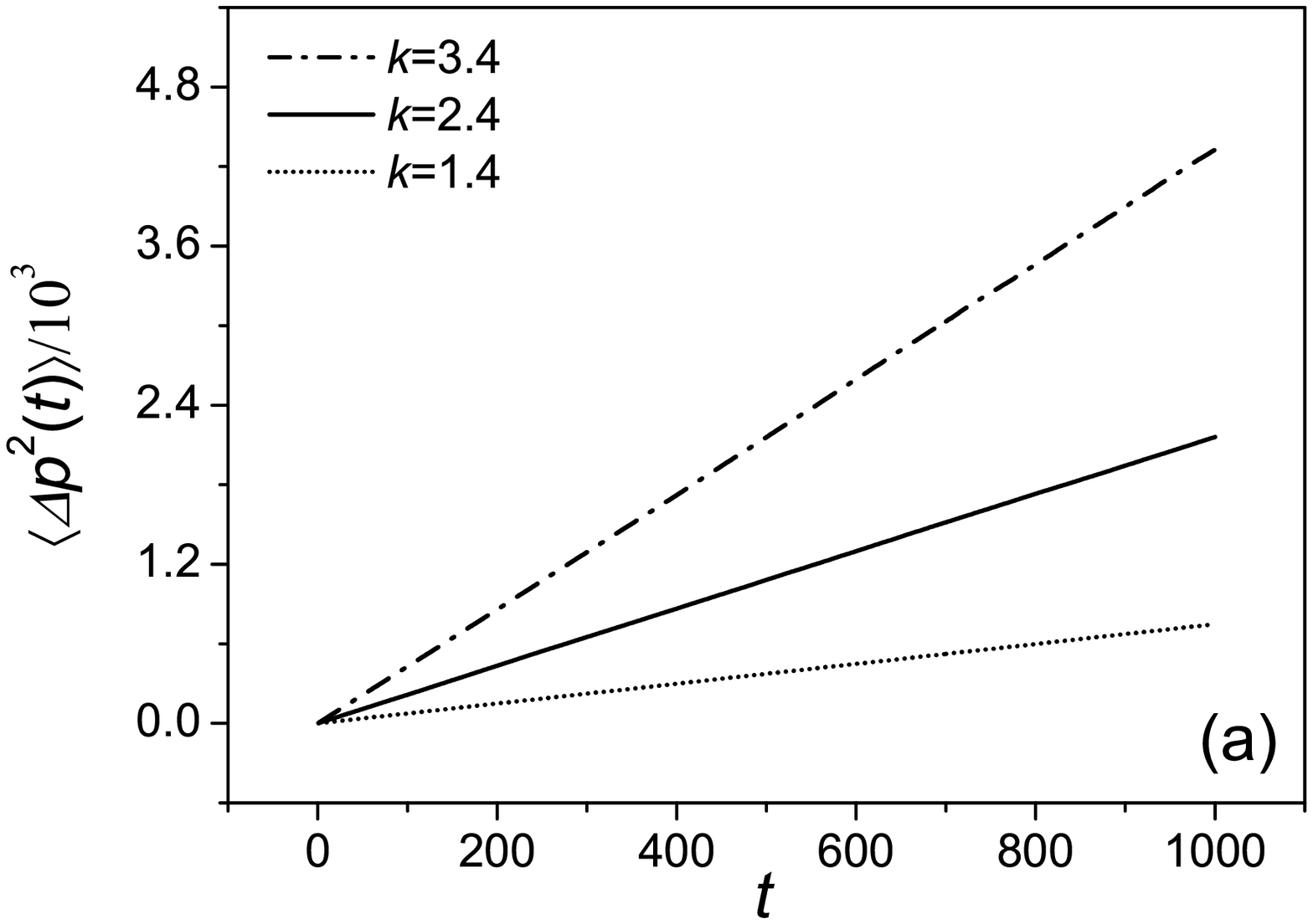}
\includegraphics[width=0.95\columnwidth,clip]{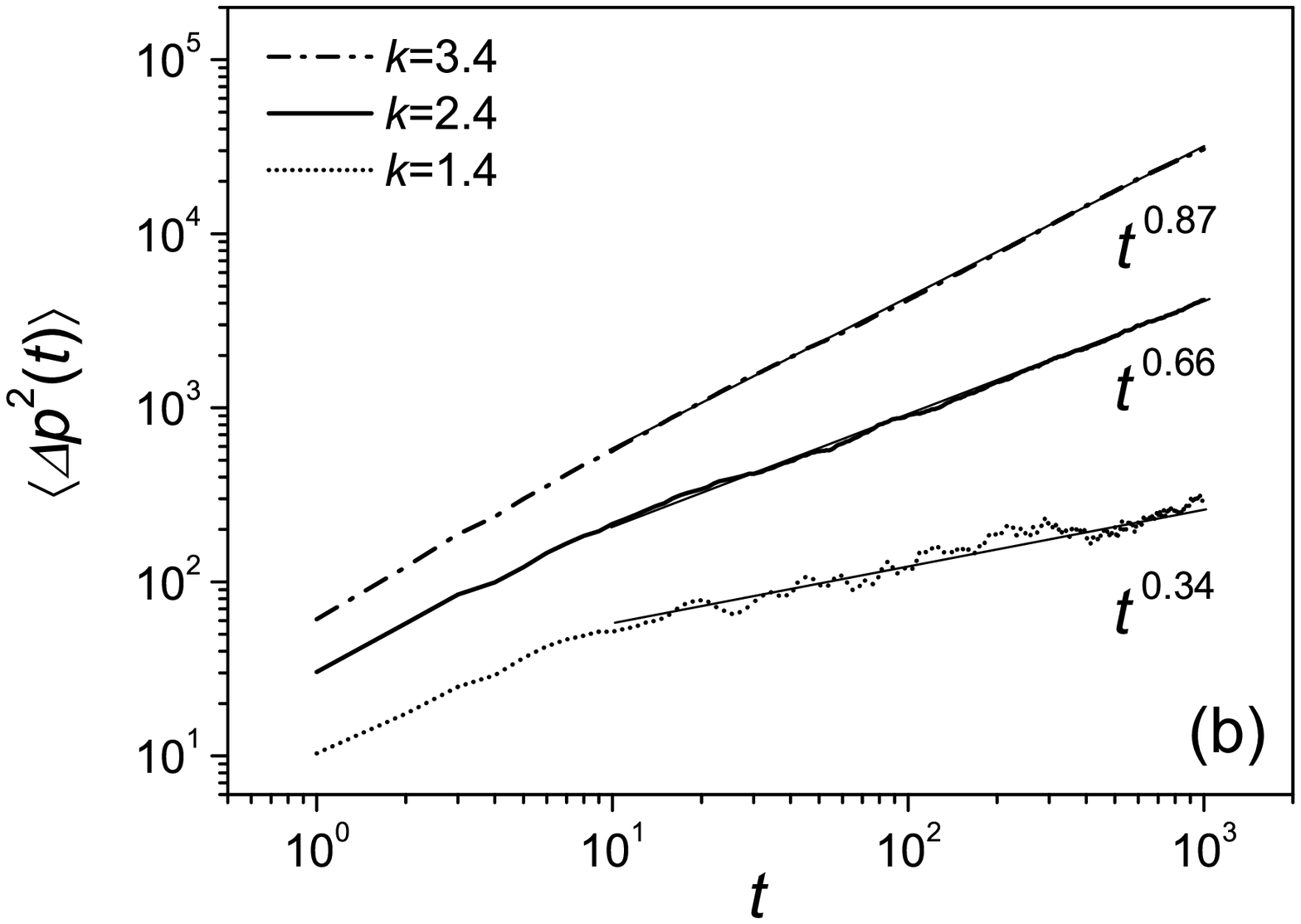}
\caption{Classical (a) and quantum (b) diffusion,  $\langle \Delta
p(t)^2 \rangle \equiv \langle [p(t)-p(0)]^2 \rangle$,  for three
different kicking values, above, below and close to the transition
$k_c \approx 2.4$ in the 3d kicked rotor Eq.(\ref{3dkr}) with
$\{\tau_i\}$ as in Fig. 1(a). Classical diffusion is normal for the
three kicking values however interference effects slows down and
eventually arrests quantum diffusion for $k < k_c$.  In agreement
with OPT $\langle p^2 \rangle \propto t^{2/3}$ at $k \approx k_c$.
Thin solid lines in (b) are the best fits in the window
$10<t<1000$.}\label{fig3}
\end{figure}
%%%%%%%%%%%%%%%%%%%%%%%%%%%%%%%%%%%%%%%%%%%%%%%%%%%%%%%%%%%%%%%%%%%%%%%%%%%%%%%%%%%%%%%%%%%%%%%%%%%fig 3

\section{Quantum diffusion}

We now study quantum diffusion in the region $k \approx k_c$. Our
motivation is to find out whether the predictions of the scaling
theory still hold in the 3d kicked rotor Eq.(\ref{3dkr}).

Provided that the classical phase is fully chaotic we expect the
classical motion to be diffusive $\langle \Delta p^2(t) \rangle
\propto t$. By contrast quantum dynamics depends strongly on $k$.
In analogy with a prediction of the OPT for the 3d AM, we expect
dynamical localization for $k < k_c$. In the opposite limit,
quantum effects are small, dynamical localization is avoided and
diffusion does not stop though it might slow down as a consequence
of interference effects.

%%%%%%%%%%%%%%%%%%%%%%%%%%%%%%%%%%%%%%%%%%%%%%%%%%%%%%%%%%%%%%%%%%%%%%%%%%%%%%%%%%%%%%%%%%%%%%%%%%%fig 4
\begin{figure}
\includegraphics[width=0.95\columnwidth,clip]{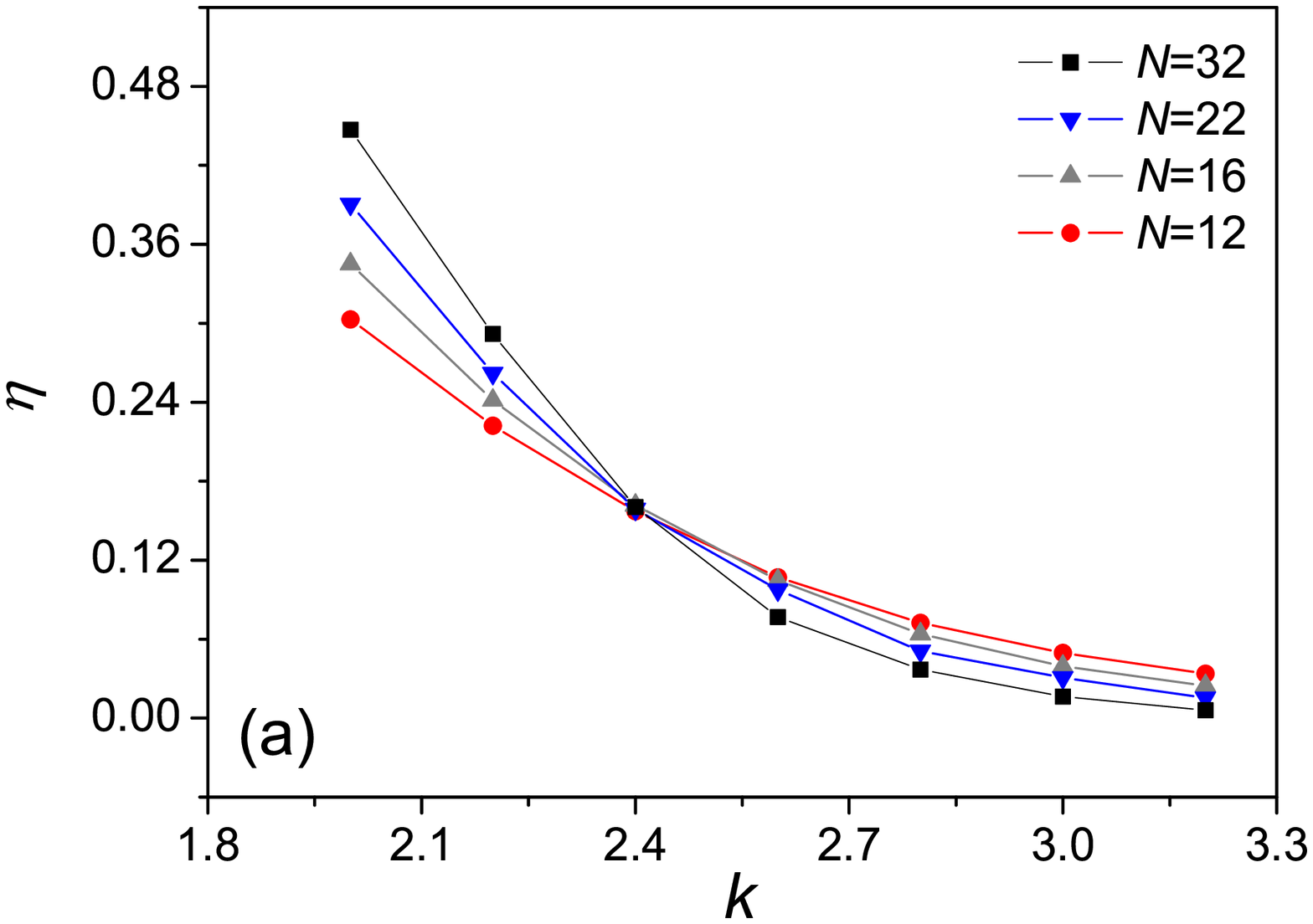}
\includegraphics[width=0.95\columnwidth,clip]{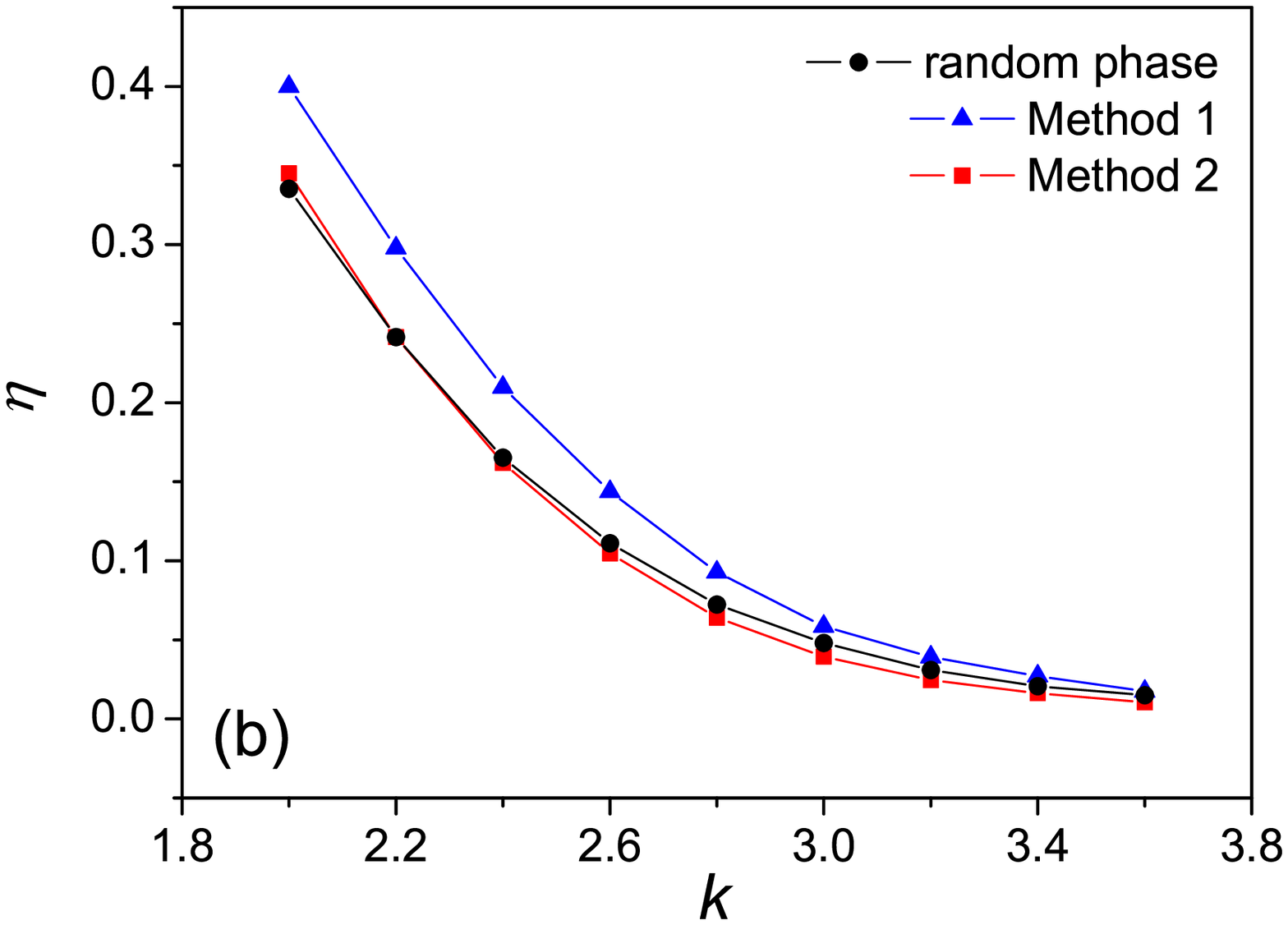}
\caption{Upper: $\eta$ versus the kicking strength for different
system sizes $N$ for a 3d kicked rotor in which the phase of the
kinetic term of the evolution matrix is random (see text). A
transition is again observed but for a slightly bigger $k_c = 2.4\pm
0.05$. Lower: $\eta$ for a randomized 3d kicked rotor (random phase
plot) and for the deterministic Eq.(\ref{3dkr}) one with
$\{\tau_i\}$ chosen from the two methods mentioned in the text. In
both cases $N=16$. Small but sizable differences are clearly
observed specially in the insulator region. For these $\{\tau_i \}$
correlations in the deterministic 3d kicked rotor seem to slightly
weaken localization effects. In order to a meaningful comparison in
the deterministic case we carry out a full ensemble average over
$1280$ sets of $\{\tau_i \}$.} \label{fig4}
\end{figure}
%%%%%%%%%%%%%%%%%%%%%%%%%%%%%%%%%%%%%%%%%%%%%%%%%%%%%%%%%%%%%%%%%%%%%%%%%%%%%%%%%%%%%%%%%%%%%%%%%%%fig 4

According to the OPT (see introduction) for $k \approx k_c$,
quantum diffusion must be anomalous $\langle \Delta p^2 (t)
\rangle \propto t^{2/3}$. In Fig. \ref{fig3}, we show the
classical and quantum $\langle \Delta  p^2(t) \rangle$ for
different values of the kicking strength. We find full agreement
with the predictions of the OPT: dynamical localization is avoided
for $k > k_c$ and, for $k \approx k_c$, diffusion is anomalous
with $\langle \Delta p^2 (t) \rangle \propto t^{\beta}$, $\beta
\approx 0.66$. We note that the determination of $k_c$ using the
finite size scaling method is fully consistent with the results
from quantum diffusion.

The classical  $\langle \Delta p^2 (t) \rangle$ was computed by
averaging over an ensemble of $10^6$ initial $(t = 0)$ states
uniformly distributed in $-\pi<q_i(0)<\pi$ with $p_i(0)=0; i=1,2,3$.
In the quantum case we average over the following four initial
conditions, $|{\bf n}\rangle=|0,0,0\rangle, |0,0,1 \rangle,
|0,1,0\rangle$ and $|1,0,0\rangle$.

\section{Effect of correlations: Comparison with a 3d kicked rotor with random kinetic term}

In this section we compare previous results with those coming from a
3d kicked rotor similar to Eq.(\ref{3dkr}) but with the kinetic term
of the evolution matrix completely random, namely
%{\bf add some more details if you find it necessary},
the factor $e^{i(\tau_1 n_1^2 + \tau_2 n_2^2 + \tau_3 n_3^2)}$ is
replaced by $e^{i\phi}$ where  $\phi$ is a random number in
$[0,2\pi)$.

Our motivation is to clarify to what extent the deterministic but
pseudo-random character of the motion of the 3d kicked rotor
Eq.(\ref{3dkr}) weakens Anderson localization effects. This is
relevant for applications as in realistic situations it is hard to
produce a truly random potential free from correlations.  In essence
we repeat the calculations of the previous section but for the
randomized version of the 3d kicked rotor: a) we carry out a finite
size scaling analysis to determine $k_c$ and $\nu$ (see Fig.
\ref{fig4}), b) we compute the spectral rigidity and level spacing
distribution in the metallic, critical and insulator region, c) we
study quantum diffusion for $k \approx k_c$.

%%%%%%%%%%%%%%%%%%%%%%%%%%%%%%%%%%%%%%%%%%%%%%%%%%%%%%%%%%%%%%%%%%%%%%%%%%%%%%%%%%%%%%%%%%%%%%%%%%%fig 5
\begin{figure*}
\includegraphics[width=.95\columnwidth,clip]{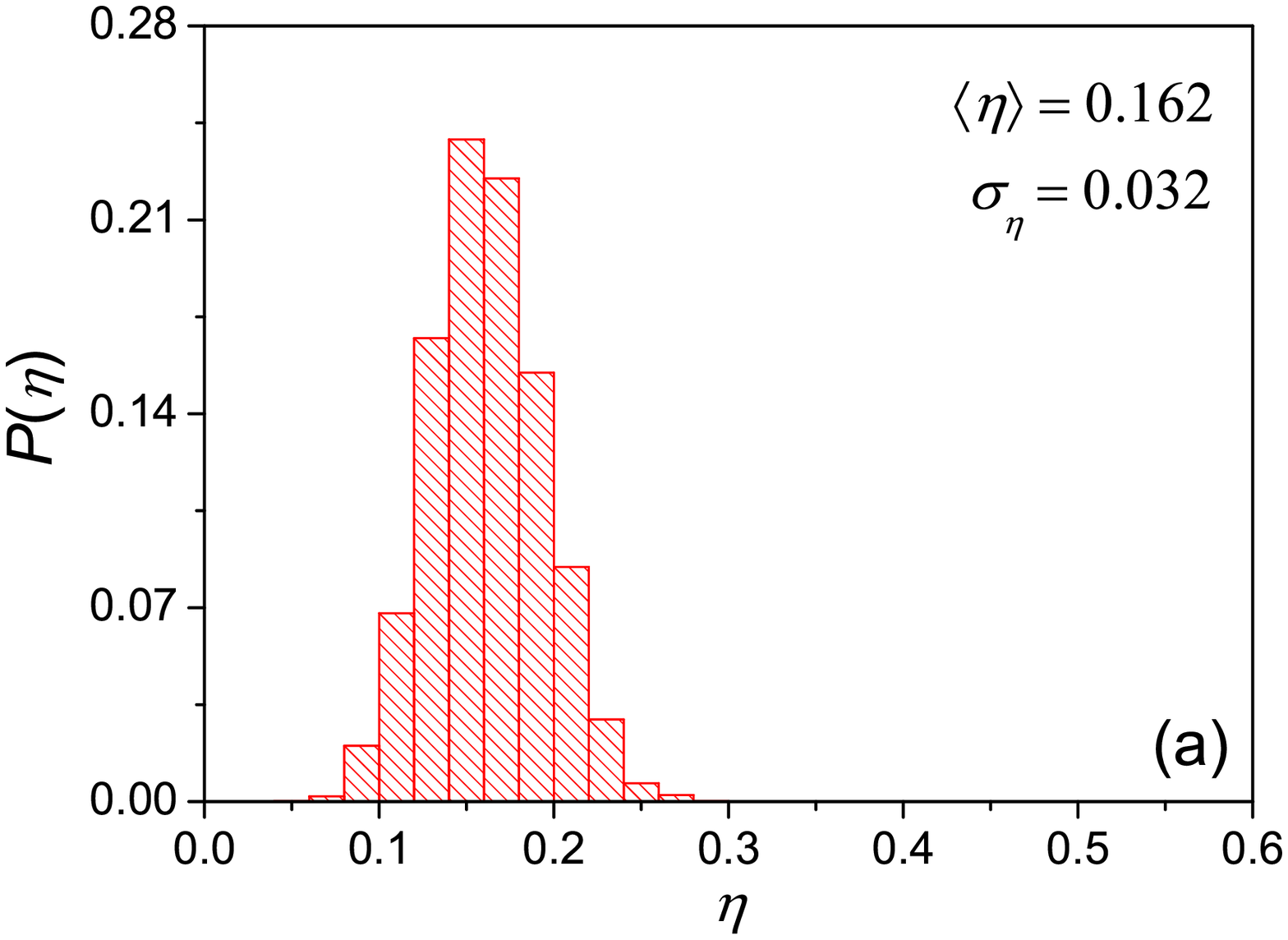}
\includegraphics[width=.95\columnwidth,clip]{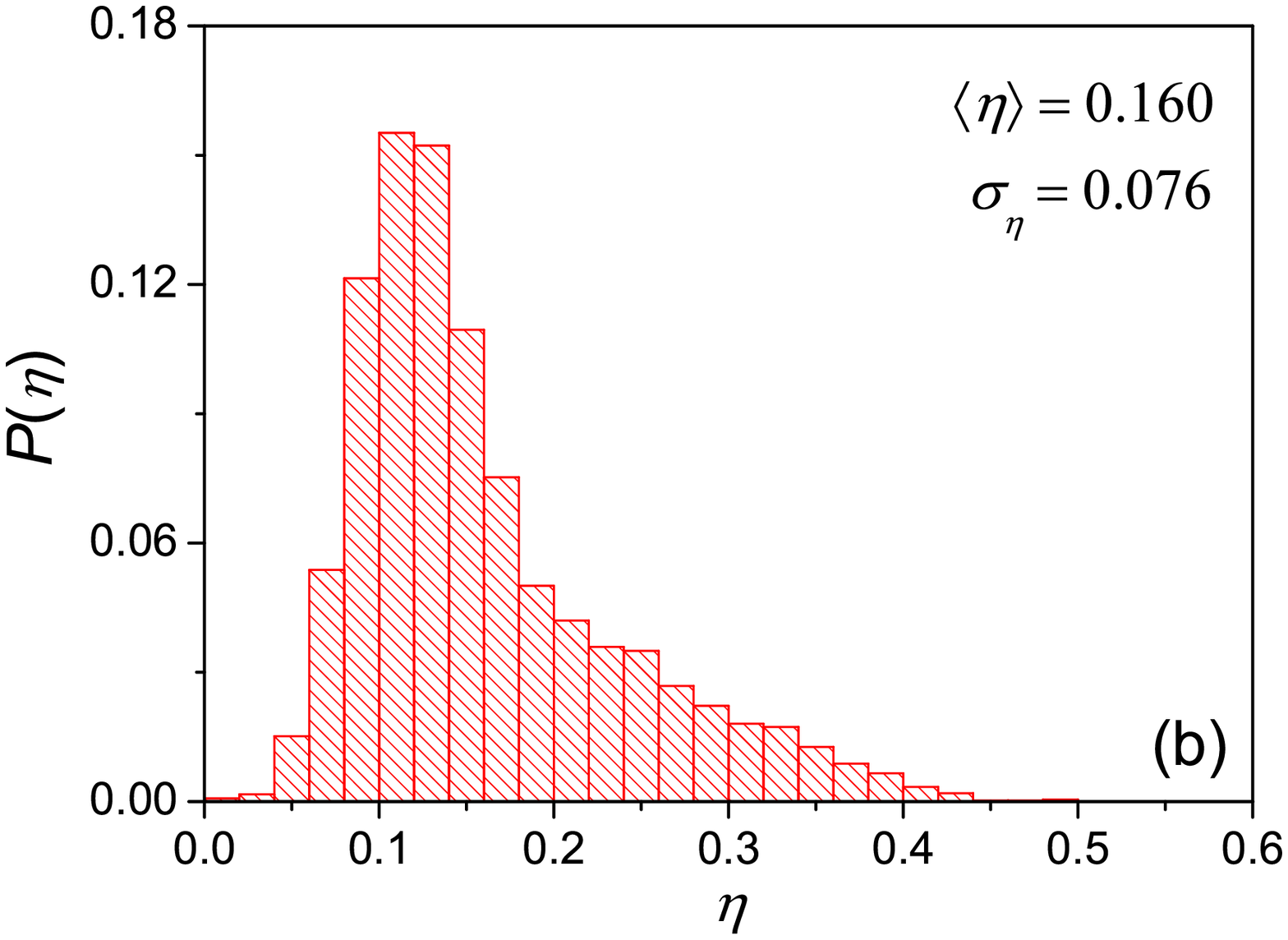}
\includegraphics[width=.95\columnwidth,clip]{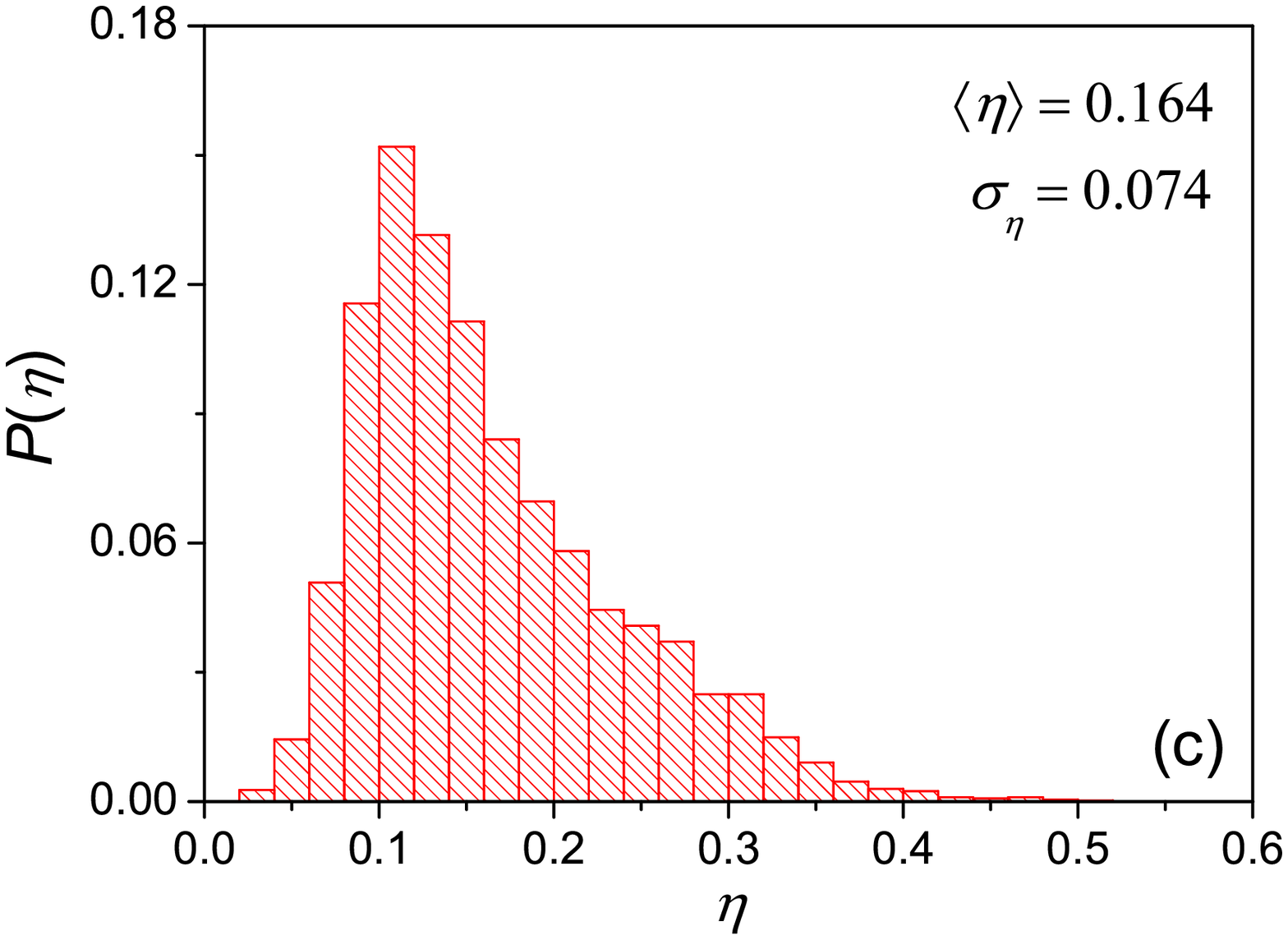}
\includegraphics[width=.95\columnwidth,clip]{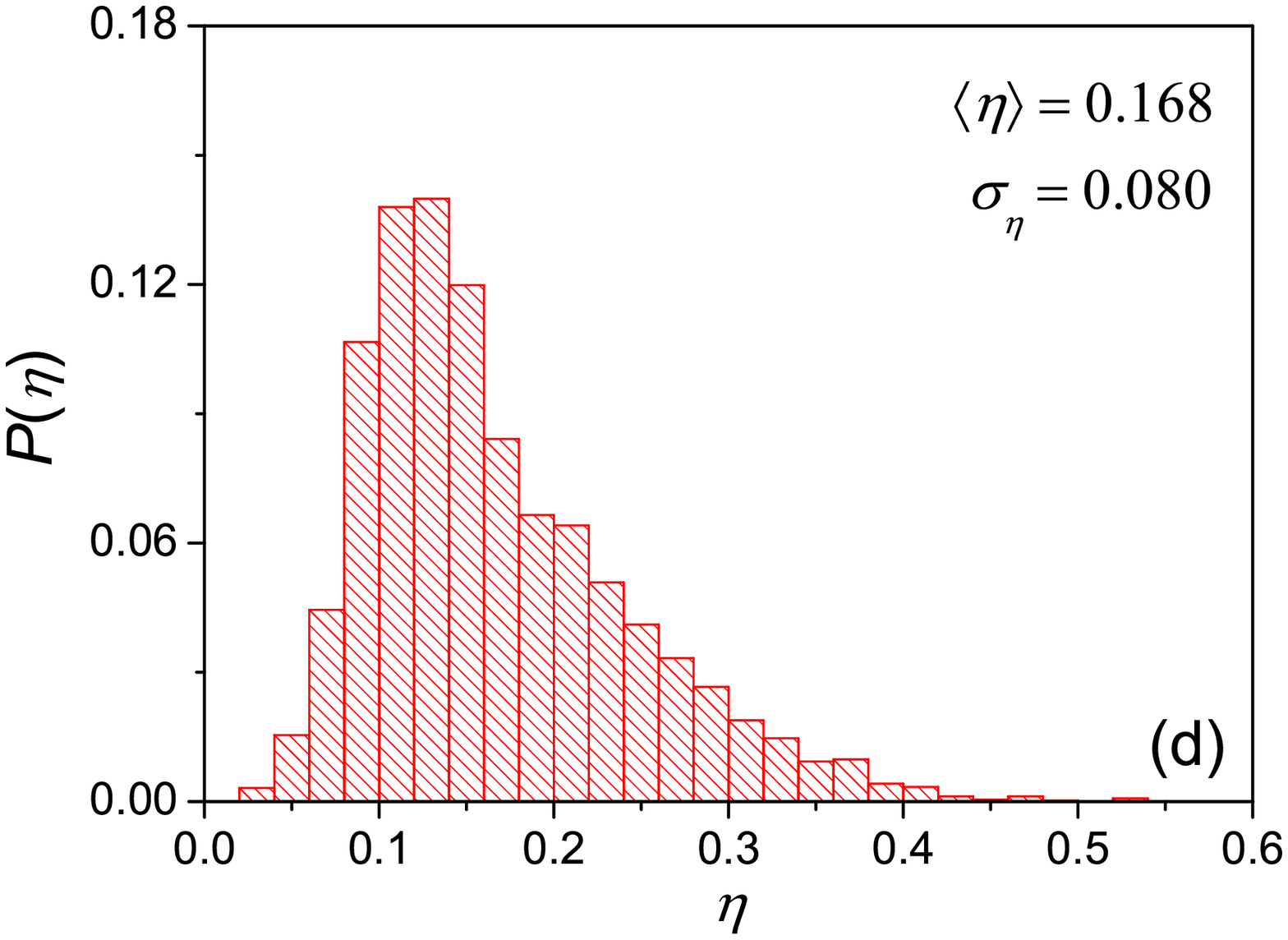}
\caption{Distribution of $\eta$ for a 3d kicked rotor
Eq.(\ref{3dkr}) with (a) the kinetic term replaced by a random
phase, (b) $\{\tau_i\}$ given by Method 1, (d) $\{\tau_i\}$ given
by Method 2, and (c) the same for Method 2 but where the integer
sequence $[a_1, a_2, ...]$ for the continuous fraction form of
$\tau_i/\pi$ satisfies $|a_i|<9$ instead. In all cases 1280 sets
of $\{\tau_i\}$ are considered and $N =16$.} \label{fig5}
\end{figure*}
%%%%%%%%%%%%%%%%%%%%%%%%%%%%%%%%%%%%%%%%%%%%%%%%%%%%%%%%%%%%%%%%%%%%%%%%%%%%%%%%%%%%%%%%%%%%%%%%%%%fig 5

The main conclusions are as follows:\\
a) A transition is also observed in this model but at a slightly
greater $k \approx k_c =2.4$ than in the deterministic case (see fig
\ref{fig4}). \\
b) The slope of the number variance  $\chi \approx 0.22$ is smaller
than in the deterministic case. This is consistent with weaker
localization effects in the deterministic case.\\ c) $\nu \approx
1.6 \pm 0.2$ is roughly the same in both cases. This is expected as
the numerical value of critical exponents should not depend on the microscopic details of the system.\\
 d)Quantum
diffusion is anomalous with $\langle p^2(t) \rangle \propto t^{2/3}$
and virtually indistinguishable from both the deterministic and the
3d AM. This together with the scale invariance of the spectral
correlations, the level repulsion, the asymptotic linear behavior of
the spectral rigidity and the numerical value of critical exponent
$\nu$ seem to be the more robust features of the transition.

A remark is in order. The results for the random case were
obtained by performing an ensemble average over $1280$ different
realizations. By contrast in the deterministic case we did not
want to perform an artificial ensemble average so the results
presented corresponds to only a handful of different $\{\tau_i
\}$.
%Finally, for the sake of completeness we show {\bf can we be more
%explicit?} in \ref{fig6} the correlation for two different type of
%irrational %numbers. Clearly these correlations depend on how we
%choose the irrational number but.Although they cannot drive the
%system away the transition can renormalize the different parameters
%that characterize the transition.
A natural question to ask  is to what extent the numerical values of
$k_c$ will be different if other $\{\tau_i \}$ would have been used.
 For
the $\{\tau_i \}$ used here we have obtained a $k_c$ which is
slightly smaller than in the random case. By contrast in a recent
paper \cite{us2008} we found $k_c \approx 3.3$ and $\eta_c \approx
0.06$ for $\tau_1 = 1$, $\tau_2 = 1/\lambda$ and $\tau_3 =
1/\lambda^2$ with lambda the root of $x^3-x-1=0$. This value of
$k_c$ is sizable larger than the one $k_c \approx 2.4$ found in the
random case. This suggests that for that choice of $\{\tau_i \}$ the
effect of correlations enhances localization effects. Several
questions then arise:  what is the probability that a given set of
periods $\{\tau_i \}$ leads to an enhancement/supression of
localization?, on average, what is the magnitude of the typical
deviations from the random case?

Unfortunately a detailed answer to these questions faces technical
difficulties as the finite size scaling method is quite expensive
numerically. Moreover, on the more fundamental level, this is rather
a number theory problem that, as far as we know, have not yet been
tackled in the mathematical literature.

In order to give a qualitatively answer to these questions we
compute the full distribution of $\eta$, its average $\langle \eta
\rangle $ and standard deviation $\sigma_{\eta}$ where $\langle
\ldots \rangle$ in this case stands for spectral and ensemble
average (see Fig. 5). In the deterministic case we carry the
ensemble average over different sets of $\{\tau_i \}$: from Method
1, Method 2 and Method 2 but with $|a_i| < 9$. In each case $1280$
different choices of $\{\tau_i \}$ are considered. In the random
case the average is carried out also over $1280$ different random
realizations as it is explained above. Several conclusions can be
extracted from this numerical calculation (see Fig. 5): a) The
distribution in the random case is much narrower than in the
deterministic case and as a consequence the standard deviation is
much smaller. It is thus expected that specific choices of $\{\tau_i
\}$ could lead to sizable deviations between the random and
deterministic case. b) The distribution of $\eta$ for the different
methods of obtaining $\{\tau_i \}$ are qualitatively similar. That
suggests that provided that is sufficiently irrational ($|a_i| <12$)
the distribution of $\eta$ is not very sensitive to the way in which
$\{\tau_i \}$ is generated. c) The distribution of $\eta$ is clearly
asymmetric. A longer tail is observed in the region of greater
$\eta$ which indicates that for a sizable percentage of $\{\tau_i
\}$ localization effects are considerably suppressed. At the same
time the maximum of the distribution is slightly shifted to the left
with respect of the random case. Therefore for a typical $\{\tau_i
\}$, $k_c$ will be somehow smaller than in the random case and
consequently localization effects will be mildly suppressed due to
the non random character of the motion.

\section{Conclusions}
We have studied the quantum dynamics of a 3d kicked rotor with a
smooth potential. A careful finite size scaling analysis has shown
that this system undergoes a metal-insulator transition for $k
\approx k_c$. The value of $k_c$ barely depends on the numerical
choice of $\{\tau_i \}$ provided it is sufficiently irrational.
However from the analysis of the distribution of $\eta$ it is likely
that for particular choices of irrational $\{\tau_i \}$ stronger
deviations are observed.

 In the critical region $k \approx k_c$
we have investigated in detail quantum diffusion, level statistics
and critical exponents. It has been shown that quantum diffusion, in
agreement with the prediction of the OPT, is quantitatively similar
to that of a 3d AM at the transition. Typical signatures of a
metal-insulator transition such as a linear spectral rigidity or
level repulsion have also been observed in our 3d kicked rotor for
both deterministic and random kinetic terms.

\acknowledgements

JW is grateful to professor C.-H. Lai for his encouragement and
support, and acknowledges support from Defence Science and
Technology Agency (DSTA) of Singapore under agreement of POD0613356.
AMG acknowledges financial support from a Marie Curie Outgoing
Action, contract MOIF-CT-2005-007300 and also from the FEDER and the
Spanish DGI through Project No. FIS2007-62238.

\end{document}